\documentclass[11pt,a4paper]{article}
\pdfoutput=1
\usepackage{jheppub}
\usepackage{amsfonts,amsbsy,amsmath,amssymb}
\usepackage{graphicx}
\usepackage{booktabs}
\usepackage{caption}
\usepackage{subcaption}
\usepackage{multirow}
\usepackage{algorithm}
\usepackage[noend]{algpseudocode}
\newcommand{\be}{\begin{equation}}
\newcommand{\ee}{\end{equation}}

\title{A comparison of updating algorithms for large $N$ reduced models}

\author[a]{Margarita Garc\'{i}a P\'erez,}
\author[a,b]{Antonio Gonz\'alez-Arroyo,}
\author[c]{Liam Keegan,}
\author[d,e]{\mbox{Masanori Okawa}}
\author[c]{and Alberto Ramos}

\affiliation[a]{Instituto de F\'{i}sica Te\'orica UAM-CSIC, Nicol\'as
  Cabrera 13-15, Universidad Aut\'onoma de Madrid, E-28049–Madrid,
  Spain}

\affiliation[b]{Departamento de F\'{i}sica Te\'orica, C-XI Universidad 
Aut\'onoma de Madrid, E-28049 Madrid, Spain}
\affiliation[c]{PH-TH, CERN, CH-1211 Geneva 23, Switzerland}
\affiliation[d]{Graduate School of Science, Hiroshima University,
  Higashi-Hiroshima, Hiroshima 739-8526, Japan}
\affiliation[e]{Core of Research for the Energetic Universe, Hiroshima University,
                     Higashi-Hiroshima, Hiroshima 739-8526, Japan}
\emailAdd{margarita.garcia@uam.es}
\emailAdd{antonio.gonzalez-arroyo@uam.es}
\emailAdd{liam.keegan@cern.ch}
\emailAdd{okawa@sci.hiroshima-u.ac.jp}
\emailAdd{alberto.ramos@cern.ch}

\abstract{
We investigate Monte Carlo updating algorithms for simulating
$SU(N)$ Yang-Mills fields on a single-site  lattice, such as for
the Twisted Eguchi--Kawai model (TEK). We show that 
performing only over--relaxation (OR) updates of the gauge links is a
valid simulation algorithm for the Fabricius and Haan formulation of this
model, and that this decorrelates observables faster than using 
heat--bath updates. We consider two different methods of implementing
the OR update: either updating the whole $SU(N)$ matrix at once, or 
iterating through $SU(2)$ subgroups of the $SU(N)$ matrix, we find the
same critical exponent in both cases, and only a 
slight difference between the two.
}

\preprint{%
{\flushright 
CERN-PH-TH-2015-030\\
IFT-UAM/CSIC-15-032\\
FTUAM-15-10\\
HUPD-1502\\
}}

\begin{document}
\maketitle

\section{Introduction}

As originally proposed by t'Hooft~\cite{'tHooft:1973jz}, the large $N$
limit of $SU(N)$ Yang-Mills (YM) theories at fixed t'Hooft coupling
is an approximation to the  model of  strong  interactions. 
Being simpler from an analytic point of view, it was hoped that it could
lead to an understanding of the properties of hadrons at low
energies. Despite the fact that only a subset of Feynman diagrams
contribute in this limit, these initial hopes turned out to be too
optimistic. Large $N$ YM theories are full of rich phenomena but 
complicated enough to resist a complete understanding. In parallel,
its interest has grown and extends much beyond its role as an
approximation to strong interactions. The t'Hooft limit  appears as an 
important ingredient in many recent developments, such as the connections 
of field theory with string theory and/or gravitational interactions 
as in the AdS/CFT construction.

Given its interest, several authors have used lattice gauge theory
techniques to study the non-perturbative behaviour of  YM theories 
in the large N limit (for a recent review see ~\cite{Lucini:2012gg}).  
This program is challenging, since the number of degrees of freedom
that have to be simulated in a  computer increases with the  rank of the
group, making the simulations  more difficult with growing $N$.
Alternatively, one can exploit the idea of volume reduction. 
The study of the Schwinger-Dyson equations of $U(N)$ YM theories led to
the conjecture that gauge theories become volume independent in the large $N$
limit~\cite{Eguchi:1982nm}. Taking the idea of volume reduction to an
extreme, one can simulate a lattice with one single point. This allows
one to reach much larger values of $N$ and can provide a more precise method of
determining the large $N$ observables. 

Attempts to produce a successful implementation of the volume
independence idea have lead to several proposals~\cite{Bhanot:1982sh,GonzalezArroyo:1982ub,GonzalezArroyo:1982hz,
GonzalezArroyo:2010ss,Kiskis:2003rd,Kovtun:2007py,Basar:2013sza,Unsal:2008ch,Azeyanagi:2010ne} 
after the original  one  was shown not to work~\cite{Bhanot:1982sh} (see also 
Ref.~\cite{Lucini:2012gg} for a  recent account). Our study here
will focus on the twisted reduction idea originally proposed by some of
the present authors~\cite{GonzalezArroyo:1982ub}, but its scope applies to other
reduced  models as well. In particular, the one-site
\emph{twisted Eguchi-Kawai model}(TEK)~\cite{GonzalezArroyo:1982hz} 
with fluxes chosen in a suitable range~\cite{GonzalezArroyo:2010ss} has been 
tested recently in several works. There is now strong numerical 
evidence that its results for various lattice observables coincide with 
those of $SU(N)$ gauge theories extrapolated to 
$N\rightarrow\infty$~\cite{Gonzalez-Arroyo:2014dua}. Furthermore, the 
tests have also extended to physical quantities in the continuum limit 
such as the string tension~\cite{GonzalezArroyo:2012fx} or  the
renormalized running coupling~\cite{Perez:2014isa}.

In all the previous works a connection was established between the 
finite $N$ corrections of the twisted model and finite volume effects
of the ordinary gauge theory. Typically $N^2$ plays the role of the
physical volume. Hence,  in order to extract physical 
quantities with small systematic errors, one should work at 
rather large values of $N$ ($\mathcal O(1000)$). These types of 
simulations present their own challenges, which are sometimes very 
different to the ones we are used to face in usual lattice simulations.
The literature is somewhat scarce and old (this will be reviewed in
the next section). This serves as a motivation for the present  work,
devoted  to the  analysis of the  computational  aspects of this one-site
model. We will present the different algorithms that can be used in the
simulations, and numerical tests of the correctness and efficiency of
these algorithms.

We must conclude by mentioning that the interest of the reduced matrix
models extends beyond their purely computational one. First of all, 
the methodology developed here is useful for extensions of pure
Yang-Mills theories such as the model with fermionic fields in the
adjoint representation with various types of boundary
conditions~\cite{Kovtun:2007py,Gonzalez-Arroyo:2013bta}. This allows the study of a possible
candidate for walking technicolor and the determination of its anomalous
dimensions~\cite{Perez:2014gqa,Perez:inprep}. There are also other intriguing phenomena, such
as a stronger form of $N$-$L$ scaling~\cite{Perez:2013dra},~\cite{Perez:2014sqa}, the emergence of new
symmetries~\cite{Basar:2013sza} or the connections with non-commutative
space-time~\cite{GonzalezArroyo:1983ac,Douglas:2001ba,Aoki:1999vr,Ambjorn:2000nb}. In summary, we are quite confident that the 
methodology studied here will be useful for future researchers in the
field.

%%% Local Variables:
%%% mode: latex
%%% TeX-master: "paper"
%%% End:

\section{Update algorithms for the one-site model}

The action of the TEK model in $d$ space-time dimensions is defined
as a function of the variables $U_\mu$ ($\mu = 0,\dots,d-1$)
that are elements of the group $SU(N)$~\cite{GonzalezArroyo:1982hz} 
\begin{equation}
  \label{eq_action}
  S_{\rm TEK}[U] = b N \sum_{\mu \ne \nu}  {\rm Tr} 
  \left( 1\!\!1 - z_{\mu \nu} U_\mu U_\nu U_\mu^\dagger U_\nu^\dagger \right)\,.
\end{equation}
where $b$ can be interpreted as the inverse of the 't Hooft coupling
$\lambda = g^2 N$, and $z_{\mu \nu}$ is given by
\begin{equation}
z_{\mu \nu} = \exp \left( {2 \pi i n_{\mu\nu}\over N}\right),\quad  n_{\nu \mu} = -n_{\mu \nu}\,,
\end{equation}
where $n_{\mu\nu}$ (called the twist tensor) are integers defined
modulo $N$. The action 
Eq.~(\ref{eq_action}) is just the Wilson action of an $SU(N)$ lattice
gauge theory defined on a one--site lattice with twisted boundary
conditions.

The partition function of this model is given by
\begin{equation}
  \label{eq:pfunc}
  \mathcal Z = \int\,\mathcal D[U]\, \exp\left\{-S_{\rm TEK}[U]\right\}\,,
\end{equation}
and in principle can be simulated with the usual Monte Carlo
techniques like, for
example, the metropolis algorithm~\cite{Okawa:1982ic}. Nevertheless the most effective
algorithms used to simulate pure gauge
theories~\cite{Creutz:1980zw,Cabibbo:1982zn,Creutz:1987xi,Adler:1987ce,Brown:1987rra,Wolff:1992nq}, a
combination of  
heatbath (HB) and overrelaxation (OR) sweeps, cannot be applied
directly to this model. The reason is that the action
Eq.~(\ref{eq_action}) is not a linear function of the links
$U_\mu$. The solution to this problem was discovered long ago by
Fabricius and Haan~\cite{Fabricius:1984wp} and consists in introducing
auxiliary normally distributed complex $N\times N$ matrices 
$\widetilde Q_{\mu\nu}$ with $\mu>\nu$. The
original partition function Eq.~(\ref{eq:pfunc}) can be written as 
\begin{equation}
  \mathcal Z = {\mathcal N} \int\,\mathcal D[U]
  \mathcal D[\widetilde Q]\,
  \exp\left\{-S_{\rm TEK}[U]\right\}\, 
  \exp\left\{-\frac{1}{2}\sum_{\mu > \nu} {\rm Tr}\left(
        \widetilde Q^\dagger_{\mu\nu} \widetilde Q_{\mu\nu}\right)\right\}
  \,,
\end{equation}
where ${\mathcal N}$ is a constant. If we perform the change
of variables 
\begin{equation}
  \label{eq:Q}
  \widetilde Q_{\mu\nu} = Q_{\mu\nu} - t_{\mu\nu}U_\mu U_\nu -
  t_{\nu\mu} U_\nu U_\mu\,,\quad t_{\mu\nu} =e^{\pi i
      n_{\mu\nu}/N}\sqrt{2Nb}\,, 
\end{equation}
the partition function Eq.~(\ref{eq:pfunc}) can be written as
\begin{equation}
  \label{eq:changev}
  \mathcal Z = \mathcal N \int\,\mathcal D[U]\mathcal D[Q]\,
  \exp\left\{-S_{\rm TEKQ}[U,Q]-\frac{1}{2}\sum_{\mu > \nu} {\rm Tr}\left(
        Q^\dagger_{\mu\nu} Q_{\mu\nu}\right)\right\}
  \,,
\end{equation}
where\footnote{Note that the Jacobian of the change of variables 
Eq.~(\ref{eq:Q}) is just one.} 
the modified action $S_{\rm TEKQ}$ given by
\begin{equation}
  \label{eq:actionQ}
  S_{\rm TEKQ}[U,Q] = - \sum_{\mu > \nu} 
  {\rm Re} {\rm Tr}\left[ Q_{\mu\nu}^\dagger \left(t_{\mu\nu}U_\mu U_\nu +
      t_{\nu\mu} U_{\nu}U_{\mu} \right) \right]
\end{equation}
is linear in the link variables $U_\mu$. 

At this stage standard algorithms like heatbath and overrelaxation can be applied to update the links. The Fabricius-Haan 
algorithm thus involves the updating of the auxiliary variables followed by the link updates.
We will describe below several updating possibilities and show that the optimal algorithm only requires 
over-relaxation updates of the links combined with the updating of the auxiliary variables.

\subsection{Update of the link variables}

The terms of the action that involve some link $U_\alpha$ are
\begin{equation}
\label{eq:onelink}
  S_\alpha[U_\alpha] = -{\rm Re}\,{\rm Tr}\left( U_\alpha H_\alpha\right)\,,
\end{equation}
where $H_\alpha$ is the sum of staples given by
\begin{equation}
  \label{eq:staples}
  H_\alpha = \sum_{\nu\ne\alpha} t_{\alpha\nu} U_\nu
  Q_{\alpha\nu}^\dagger +
  t_{\nu\alpha} Q_{\alpha\nu}^\dagger U_\nu\,, 
\end{equation}
and we have defined $Q_{\alpha\nu} \equiv Q_{\nu \alpha}$ for $\alpha < \nu$.

Each link $U_\alpha$ can be updated using a combination of heatbath and
overrelaxation updates. The heatbath updates are performed by
projecting into $SU(2)$ subgroups as 
described in~\cite{Creutz:1980zw,Cabibbo:1982zn,Fabricius:1984wp}. The
overrelaxion updates can be implemented 
through sequential SU(2) 
updates~\cite{Creutz:1987xi,Adler:1987ce,Brown:1987rra,Wolff:1992nq}
or by performing an overrelaxation
update of the whole $SU(N)$ 
matrix~\cite{Kiskis:2003rd,deForcrand:2005xr}. 

\subsubsection{$SU(2)$ projections}

We introduce  $SU(2)$ subgroups $\mathcal H_{(i,j)} \subset SU(N)$  
labelled by two  integers $(i,j)$ such that $i\ne j$. An
element $A$ of the subgroup $H_{(i,j)}$ can be written as
\begin{equation}
  \label{eq:Asub}
  A \in H_{(i,j)} \Longrightarrow 
  A_{kl} = \left\{
    \begin{array}{cl}
    \delta_{kl} & k,l\ne i, j\\
    a_{kl} & \text{otherwise}
\end{array}
\right.\,,
\end{equation}
with the $2\times 2$ submatrix $a_{kl}$ being an $SU(2)$ element.

The update of an $SU(N)$ matrix proceeds by choosing
a set $\mathcal S$ of these $SU(2)$ subgroups of
$SU(N)$~\cite{Okawa:1982ic,Cabibbo:1982zn} 
\begin{equation}
  \mathcal S = \left\{ \mathcal H_1,\dots,\mathcal H_k\right\}
\end{equation}
such that no subset of $SU(N)$ remains invariant under left
multiplication by $\mathcal S$. This requirement
can be satisfied with $k=N-1$ as follows
\begin{equation}
  \mathcal S_{\rm minimal} = \left\{ \mathcal H_{(i,i+1)}  
  \right\}_{i=1,\dots,N-1}\,,
\end{equation}
However covering the full group requires order $N$
steps of this kind. Hence, as experience shows,  it is more effective
to choose 
$\mathcal S$ to be composed of all the $N(N-1)/2$ subgroups of this type
\begin{equation}
  \label{eq:sets}
  \mathcal S = \left\{ \mathcal H_{(i,j)}  \right\}_{i< j} \,.
\end{equation}

If $A\in\mathcal H_{(i,j)}$ is a matrix of the type
Eq.~(\ref{eq:Asub}), it is easy to evaluate the one-link action
Eq.~(\ref{eq:onelink}) 
\begin{equation}
  \label{eq:su2p}
  S_\alpha[AU_\alpha] = - {\rm Tr}\left( a  w\right) +
  \text{terms independent of } a\,.
\end{equation}
Here $a$ and $w/|w|$ are $SU(2)$ matrices with $w= Re(\tilde w_0) + i Re(\tilde w_i) \tau_i$
obtained from the $2\times 2$ complex matrix given by 
\begin{equation}
  \tilde w \equiv \tilde w_0 + i \tilde w_i \tau_i =
\left(
    \begin{array}{cc}
\tilde W_{ii} & \tilde W_{ij} \\
\tilde W_{ji} & \tilde W_{jj} 
\end{array}
\right),\quad
\tilde W = U_\alpha H_\alpha\,.
\end{equation}

A heatbath update~\cite{Fabricius:1984wp} generates the matrix $a$
with probability $e^{-S_\alpha}$. An overrelaxation 
update generates $a$ by reflecting around the matrix $w$ 
\begin{equation}
\label{eq:or2}
  a=(w^\dagger)^2/|w|^2\,.
\end{equation}
In both cases the matrix $A$ is then used to update the link $U_\alpha$
\begin{equation}
  U_\alpha \rightarrow A U_\alpha\,.
\end{equation}
We will call a sweep an update for
each of the links by all of the $SU(2)$ subgroups in
$\mathcal S$ (Eq.~(\ref{eq:sets})). A sweep therefore consists of the
left multiplication of the link $U_\alpha$ by $k=N(N-1)/2$ matrices of
the form Eq.~(\ref{eq:Asub}) for each $\alpha = 0,\dots,d-1$.

It is important to note that an overrelaxation update does not change
the value of  
 Eq.~(\ref{eq:su2p}), and hence the value of the action $S_{\rm  TEKQ}[U,Q]$ remains invariant.  

\subsubsection{U(N) overrelaxation}

The authors of~\cite{Kiskis:2003rd} proposed to perform an
overrelaxation update of the whole $SU(N)$ matrix at once. The starting
point is again the link $U_{\alpha}$ which we want to update according to
the probability density
\begin{equation}
  \label{eq:prob}
  P(U_{\alpha}) \propto \exp\left(-{\rm ReTr} \left[ U_\alpha H_\alpha \right]\right)\,.
\end{equation}
where $H_\alpha$ is the sum of staples defined in Eq.~(\ref{eq:staples}).
The $SU(N)$ overrelaxation step  consists of a proposed update of the form
\begin{equation}
\label{eq:orn}
U^{\rm new}_{\alpha} = W_{\alpha} U_{\alpha}^{\dagger} W_{\alpha}\,.
\end{equation}
where  $W_{\alpha}$ is an SU(N) matrix that does not depend on 
$U_{\alpha}$ or $U^{\rm new}_{\alpha}$. The change is accepted or
rejected with probability $P(U^{\rm   new}_{\alpha})/P(U_{\alpha})$.
This is a valid update algorithm, but its efficiency depends on the
acceptance rate. To maximize it one should  minimize the change in 
the action caused by the proposed update. In fact, as we will see,
if the link variables were  elements of the U(N) group, the procedure described 
in~\cite{Kiskis:2003rd,deForcrand:2005xr} results in an exact microcanonical
update (i.e. the action $S_{\rm   TEKQ}[U,Q]$ is unchanged by the update).
The construction is as follows. 
Given the singular value decomposition of the staple
\begin{equation}
H_{\alpha}=X_{\alpha} \sigma V^{\dagger}_{\alpha}\,,
\end{equation}
where $\sigma$ is diagonal and real, and $X,V$ are $U(N)$ matrices,
we construct the matrix
\begin{equation}
W_{\alpha}=V_{\alpha} X^{\dagger}_{\alpha}\,.
\end{equation}
This matrix is an element of $U(N)$ by construction, and it is a
straightforward exercise to check that the action $S_{\rm TEKQ}[U,Q]$ is
left unchanged by the transformation of Eq.~(\ref{eq:orn}). All
updates referred to as ORN in the rest of this paper use this method.

The algorithm is applicable to one-site models directly because for
them  there is no difference between the $U(N)$ and $SU(N)$ gauge
models. This is obvious since any possible phase factor $e^{\imath
  \alpha_\mu}$ of the links $U_\mu$ cancels in the evaluation of the
action Eq.~(\ref{eq_action}). Moreover if the observables of interest
are center-invariant (i.e. any Wilson loop-like quantity), the phases
also do not play any role in the evaluation of these observables.
This statement can be made explicit in the partition function by
introducing another set of real variables $\alpha_\mu$ uniformly
distributed in $(0,2\pi)$ and writing the probability measure as
\begin{equation}
\frac{1}{(2\pi)^d} \prod_\mu d\alpha_\mu
\end{equation}
After the change of variables $U_\mu \rightarrow e^{\imath
  \alpha_\mu}U_\mu$, the action of the $SU(N)$ TEK model transforms
into that of the $U(N)$ TEK model. We stress that this is a particularity of the one-site model, and in
that in general $SU(N)$ and $U(N)$
gauge theories on the lattice differ by a $\mathcal O(1/N^2)$
contribution.  

Given that here we are concentrating on one-site lattice models we will 
stick to the previously defined microcanonical U(N) overrelaxation
update for the rest of this paper. However, for models in which not
all directions are fully reduced the equivalence of SU(N) and U(N) 
models does not apply. For the sake of the readers we will develop upon  
the $SU(N)$ case in the appendix~\ref{ap:orsun}.

%%% Local Variables:
%%% mode: latex
%%% TeX-master: "paper"
%%% End:

\subsection{A new update algorithm for the one-site model}

Monte Carlo algorithms for these type of models consists 
of two elements: the update of the auxiliary $Q_{\mu\nu}$ matrices
followed by any combination of the possible updating procedures described in the 
previous section: heat-bath (HB), SU(2) overrelaxation steps (OR2) and
U(N) overrelaxation steps (ORN). 

The new update algorithm that we will propose and analyze numerically
in this paper can
be written in a few lines (algorithm~\ref{alg} with \emph{only}
over-relaxation updates). It requires generating $d(d-1)$
auxiliary matrices per sweep combined with only overrelaxation updates
for the link update.
Despite the absence of any HB update this is a valid algorithm since 
it satisfies detailed balance and ergodicity. The latter is proven in
the next subsection.
\begin{algorithm}[t]
\caption{Update algorithm for one site models.}
\label{alg}
\begin{algorithmic}[1]
  \For{$\alpha=0, d-1$}
    \State Generate $d-1$ auxiliary variables $Q_{\alpha\mu}$
    according to Eq.~(\ref{eq:Q})
      \State Compute the Staples $H_\alpha$ according to
        Eq.~(\ref{eq:staples})
	  \State Update $U_\alpha$ using HB/OR2/ORN
	    \EndFor
	    \end{algorithmic}
	    \end{algorithm}

Since in the one-site model the
over-relaxation updates are applied to the action $S_{\rm TEKQ}[U,Q]$,
it is the value of this action that is unchanged by over-relaxation
updates, but the original action of the TEK model $S_{\rm TEK}[U]$
does change by the combination of introducing the auxiliary variables
and performing an over-relaxation sweep. The situation has some
similarities with the HMC algorithm, where one also introduces
auxiliary variables (the random momenta), and performs an update that
leaves the Hamiltonian unchanged.

\subsection{Ergodicity of over-relaxation updates}

Here we will prove  that the generation of the auxiliary
variables followed by an overrelaxation update allows to 
reach the full space of unitary matrices with non-zero probability. 
The key ingredient in the proof is that the mapping from the auxiliary 
matrices to the staples $H_\alpha$, Eq.~(\ref{eq:staples}), is a
surjective map from the space of complex $N\times N$ matrices onto
itself, except in some exceptional situations. If we consider one 
particular direction $\alpha$, we have 
\begin{equation}
    H_\alpha = \sum_{\nu\ne\alpha} t_{\alpha\nu} U_\nu
  Q_{\alpha\nu}^\dagger +
  t_{\nu \alpha} Q_{\alpha\nu}^\dagger U_\nu\,,
\end{equation}
which is a linear map of vector spaces. To show surjectivity it is
enough to consider only one term $\nu$ in the previous sum. Dropping 
indices for that case, the transformation becomes
\begin{equation}
  H = t U Q^\dagger +
    t^* Q^\dagger U\,.
\end{equation}
where $t^*/t = z = \exp(2\pi\imath n/N)$. Surjectivity amounts to
invertibility of this transformation, which implies that the kernel 
should vanish. By means of invertible transformations this problem 
maps onto the invertibility of the following transformation:
\begin{equation}
  Q' = P(Q) = Q + zU^\dagger QU
  \end{equation}
This can be easily shown using the basis in which $U$
  is diagonal with eigenvalues $e^{\imath \delta_a}$, where the previous
  expression reads
  \begin{equation}
    Q'_{ab} = \left[ 1 + z e^{-\imath (\delta_a-\delta_b)}\right]
    Q_{ab}\,,
    \end{equation}
Thus, the map has a well defined inverse whenever the expression in
parenthesis does not vanish for any values of the indices $a$,$b$.
Vanishing  can only occur if $z=-1$ (since in
      this case $Q'_{aa}=0$) or in a set of zero measure 
      (i.e. when two different eigenvalues of $U$ obey
      some special relation). So that, except for the $z=-1$ case which 
 will be commented later, the transformation can reach an arbitrary 
 $N\times N$ complex staple  matrix.

Let us summarize. We first introduced the auxiliary
variables $\widetilde Q_{\mu\nu}$ with Gaussian probability. Then we
performed a shift to obtain the $Q_{\mu\nu}$ variables. Thus, all 
sets of non-zero measure will have a non vanishing probability. By the 
previous proof we showed that, except for the exceptional cases
mentioned earlier, this will induce a probability distribution in
staple space given by Borel and strictly positive
measure (i.e. every non-empty open subset on the space of $N\times N$
matrices has a positive measure). The next step, explained below, 
will be to show  that both OR2 and ORN updates can then produce an
arbitrary $U(N)$  matrix with a non-zero probability.

\subsubsection{ORN updates}

Let us first study
the case of an ORN update given by Eq.~(\ref{eq:orn}). In this case we
perform the SVD 
decomposition of the staple matrix
\begin{equation}
  H = X\sigma V^{\dagger} 
\end{equation}
and update $U$ according to
\begin{equation}
  U \rightarrow W U^\dagger W \quad (W = VX^\dagger)\,.
\end{equation}
We only need to show that we can generate any $W$ with non-zero
probability. This is easily seen recalling that we can write
\begin{equation}
  H^\dagger H = (V\sigma V^\dagger)^2 \,,
\end{equation}
and except in a set of zero measure the matrix $H$ is
invertible. Now the map 
\begin{eqnarray}
  \nonumber
  GL(N,\mathbb C) &\rightarrow& U(N) \\
\label{eq:maporn}
  H &\rightarrow& W^{\dagger} = H/\sqrt{H^\dagger H}\,,
\end{eqnarray}
is continuous and therefore the  
inverse map sends an open neighborhood of $W\in U(N)$
into an open neighborhood of $H\in GL(N,\mathbb C)$. Since the measure
on $GL(N,\mathbb C)$ was strictly positive (i.e. any non-empty 
open set has measure bigger than zero) we have a non-zero
probability of generating any $U(N)$ matrix.

\subsubsection{OR2 updates}

An OR2 update consists in successive left-multiplication of the link
matrix $U$ by matrices $A_{(i,j)}\in \mathcal H_{(i,j)}$. First
let us show that the matrix $A_{(i,j)}$ is an arbitrary matrix of
$\mathcal H_{(i,j)}$. If we call $P_{(i,j)}$ the projectors onto the
subspace $\mathcal H_{(i,j)}$, we have
\begin{equation}
  A_{(i,j)} = P_{(i,j)} UH P_{(i,j)}\,,
\end{equation}
and since $H$ is arbitrary, so is $UH$ and therefore
$A_{(i,j)}$ is an arbitrary element of $\mathcal H_{(i,j)}$. The
full update is given by successive multiplications
\begin{equation}
  A = \prod_{i<j} A_{(i,j)}\,.
\end{equation}
Since both the multiplication and the projection to $SU(2)$ are
continuous, the application 
\begin{equation}
  H \rightarrow A\,,
\end{equation}
is continuous. Moreover we cover all $SU(2)$ subgroups of $SU(N)$, and the
map is surjective. Therefore the inverse image of  an open
neighborhood of $W\in U(N)$ is an open non-empty subset on the space
on matrices and again  we have a non-zero probability of generating
any $SU(N)$ matrix.

\subsubsection{The singular case and partially reduced lattices}

In a typical simulation of the TEK model, the singular case ($z=-1$)
can only happen in a very particular situation, since in the TEK
simulations the rank of the group matrices is usually taken a perfect
square $N=L^2$ and the twist tensor  $|n_{\mu\nu}| = kL$ with $k$ and $L$ co-prime. 
It is easy to see that these conditions imply that our singular case can only happen with
$N=4, k=1$. 

Nevertheless these are \emph{sufficient} conditions for the algorithm
to be ergodic, but \emph{not necessary}. In fact the previous proof
shows something much stronger than ergodicity: that with a single
sweep we can go from any configuration to any other with non-zero
probability. We have performed some extensive
simulations of the worst case ($N=4, k=1$) with $\mathcal O(10^6)$
measurements, and found that both heatbath and over-relaxation
thermalize to the same values starting both from a cold or hot
configuration, and expectation values are consistent within errors.
Moreover, we have not observed any significant dependence
of the autocorrelation time with the value of $z_{\mu\nu}$ that could
indicate a loss of ergodicity. 

In any case, if the reader is interested in simulating one of the exceptional
situations one can simply perform a heatbath sweep from time to time
to mathematically guarantee the correctness of the algorithm even in
the singular case.

We also want to point out that the above proof of ergodicity also applies
to lattices in which at least two directions are
reduced~\cite{Vairinhos:2010ha}.  In this case the auxiliary variables
are introduced only for the reduced directions. Since 
at least one term in the computation of the staples
will have a contribution coming from the auxiliary variables, the
staples will 
also be arbitrary in this case, and our proof applies. The numerical
study of this case will not be covered in this work. 

\subsection{Frequency of the update for the auxiliary variables}

In this subsection we will consider possible alternatives to algorithm~\ref{alg}
based on varying the relative ratio between the frequency at which the 
auxiliary variables are generated relative to the number of link updates 
per sweep. For the latter one can use either heat-bath (HB) or
over-relaxation (OR2 or ORN) steps. In this alternative approach, 
we also alter the order in which generation and updates are performed. 
Thus, in this version (see algorithm~\ref{alg2}) one generates the 
full set of $(d(d-1)/2)$ matrices $Q_{\mu\nu}$, and then  performs 
$n$ link updates using any of the alternatives. The ratio of $Q$
generations to link updates now becomes $d(d-1)/2n$ instead of the $d(d-1)$
of algorithm~\ref{alg}. 

It should be mentioned that our proof of ergodicity for overrelaxation 
updates does not directly apply to these new algorithms. Here one cannot 
separate the problem into independent directions and has to consider 
the full linear map from the vector space of $d(d-1)/2$ auxiliary $Q$
matrices to the vector space of $d$ staples.  Notice, however, that 
for the algorithm not to be ergodic, the map from a configuration to
another must be singular (i.e. with an almost anywhere vanishing Jacobian). 
Taking into account that these maps are perfectly regular (i.e. look at
Eq.~(\ref{eq:maporn})), and that for $d>2$ there are more auxiliary
matrices than links, we think that it is quite plausible that  
algorithm~\ref{alg2} with only OR updates is ergodic as well. This is
also supported by the 
numerical evidence that we have. A formal proof might not be hard to
find, but given our preference for algorithm~\ref{alg}, based on the
results given below, we did not make the effort to include it in the
paper. 

In conclusion, our comparison of the two alternative algorithms will be 
based on the performance analysis that will follow.

To make a comparison, the first thing to examine is the time needed to update
the auxiliary variables. Generating each auxiliary variable $Q_{\mu\nu}$ requires the
generation of the random matrix $\widetilde Q_{\mu\nu}$ and two matrix
multiplications (see Eq.(\ref{eq:Q})). Since generating the random
numbers requires $\mathcal O(N^2)$ operations, while matrix
multiplication requires $\mathcal O(N^3)$ operations, we will
neglect the time needed to generate the variables $\widetilde Q_{\mu\nu}$
\footnote{This is a good approximation for the values of N we consider. For example,
  when $N>400$ the difference between using different random number generators (RNG) 
  becomes negligible even when one RNG is 2-5 times
  faster than the other.}.
On the other hand the computation of the staples attached to one link
requires $2(d-1)$ matrix multiplications in $d$ dimensions. In particular
for $d=4$, each update of the auxiliary variables requires one third
of the time required to compute the staples. Moreover all of the
previously described algorithms also require $\mathcal O(N^3)$
operations. In practical situations, we have measured that computing
the staples takes about the same \texttt{CPU} time as an update sweep.

Although,  the ratio of generations over link updates per
sweep  $d(d-1)/2n$  is smaller for the new algorithms,  from the
numerical point of view the gain is only marginal,
since the generation of the auxiliary variables takes $\mathcal
O(10\%)$ of the computer time of a typical update.

\begin{algorithm}[t]
\caption{Alternative update algorithm for one site models. As is
  discussed in the text, keeping the auxiliary variables for many
  updates results in a worse performance.}
\label{alg2}
\begin{algorithmic}[1]
  \State Generate $d(d-1)/2$ auxiliary variables $Q_{\mu\nu}$
  according to Eq.~(\ref{eq:Q})
  \For{$i=1, n$}
  \For{$\alpha=0, d-1$}
  \State Compute the Staples $H_\alpha$ according to
  Eq.~(\ref{eq:staples})
  \State Update $U_\alpha$ using HB/OR2/ORN
  \EndFor
  \EndFor
\end{algorithmic}
\end{algorithm}

\begin{figure}
\begin{center}
\includegraphics[angle=0,width=10cm]{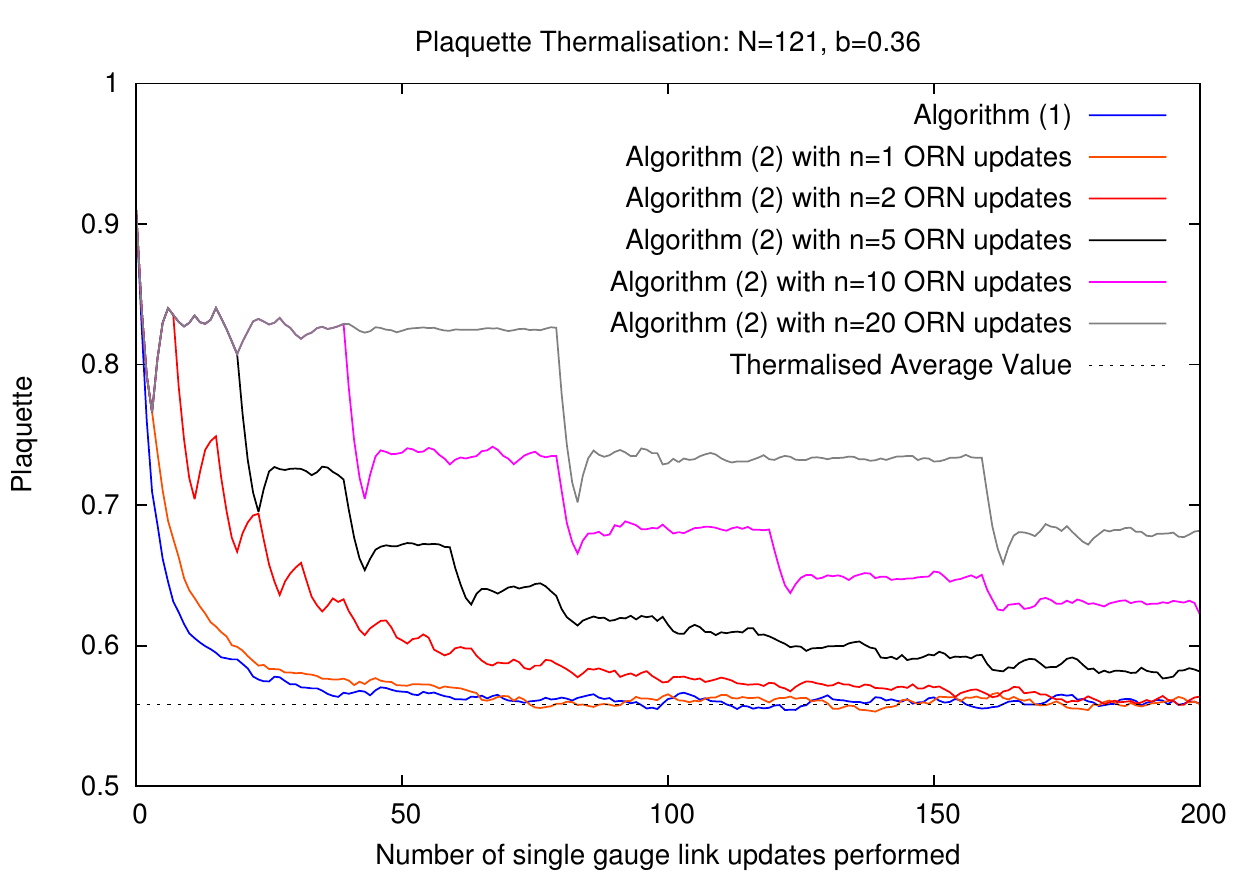}
\caption{\label{fig:therm}Comparison of the thermalization of the plaquette for $N=121$, $b=0.36, k=3$ 
 for different frequencies of $Q$--updates. It is clearly beneficial during thermalization
  to update the auxiliary valiables $Q$ as frequently as possible.}
\end{center}
\end{figure}

If from a practical point of view this approach results in
a better algorithm is a more complicated question, where
autocorrelation times have to be taken into account. Clearly the update of  
the auxiliary variables is crucial to achieve ergodicity, and
therefore if one does not 
update these variables frequently enough one might end up exploring a
small region of the space of configurations. This is nicely illustrated
by looking at the thermalization process (Figure~\ref{fig:therm}). As
the reader can see, the expectation value seems to ``plateau'' very fast
in between the updates of the auxiliary variables. Note that if one
looks at the combination of introducing the auxiliary variables and
$n$ updates, the algorithm is actually thermalizing to the correct
value, and there is no loss of ergodicity. But the figure 
indicates that performing $n>1$ updates in between the introduction of
the new auxiliary fields is redundant and does not improve the
thermalization at all.

Since the Markov operator is the same during
thermalization as during equilibrium updates, this suggests 
that autocorrelations might be significantly reduced only by the first
update after introducing the auxiliary fields. This is in fact the
case, as the data of Table~\ref{tab:wloop121Q} shows. We can see that
the most efficient algorithm is our original proposal
(algorithm~\ref{alg}), and that keeping the same set of auxiliary
variables does not help in decorrelating the measurements. 

\begin{table}[htb]
\begin{center}
\begin{tabular}{l|c|c}
      \textbf{Update algorithm} & $\langle W(1,1)\rangle$ & $\tau_{int}$ \\ \hline 
Algorithm~(\ref{alg}) & 0.558093(26) & 6.3(3) \\
Algorithm~(\ref{alg2}) with $n=1$ ORN updates & 0.558079(28) & 7.3(4) \\
Algorithm~(\ref{alg2}) with $n=2$ ORN updates & 0.558044(30) & 8.5(5) \\
Algorithm~(\ref{alg2}) with $n=5$ ORN updates & 0.558068(36) & 12.4(1.4)
  \end{tabular}
  \end{center} 
 \caption{1x1 Wilson Loop average values and integrated correlation times
   for $N=121$, $b=0.36, k=3$ with $2.5\times10^5$ sweeps, for
   different frequencies of $Q$--updates. It is clear that doing more
   ORN updates at fixed $Q$ in fact makes the autocorrelations
   worse. The integrated autocorrelation time is defined in
   Eq.(\ref{eq:tauint})}   
\label{tab:wloop121Q}
\end{table}

\subsection{Simulation algorithms and vectorization}

Let us finally comment on a few points of the previously described
algorithms. Lattice simulations usually make use of massive parallel
machines, but the simulation of one-site models is also challenging in
this respect. The link variables $U_\alpha$ are dense matrices, and a
distributed approach seems unfeasible. Nevertheless one can make use
of the vector or multi-core structure of the CPU :
\begin{description}
\item[$SU(2)$ Projections] This update allows an easy
  vectorization. Some of the $N(N-1)/2$ subgroups of the set $\mathcal
  S$ of Eq.~(\ref{eq:sets}) can be updated in parallel. For odd
  $N$, there are $l = (N-1)/2$ $SU(2)$ subgroups $\mathcal
  H_{(i,j)}$ that can be updated simultaneously. In the case of even
  $N$, there are $l = N/2$ subgroups that can
  be manipulated simultaneously. For example, for $N=5$, we have the
  following 5 sets of two subgroups that can be updated
  simultaneously:
  \begin{equation}
    \{\mathcal H_{(1, 2)}, \mathcal H_{(3, 4)}\},
    \{\mathcal H_{(1, 3)}, \mathcal H_{(2,5)}\}, \{\mathcal H_{(1, 4)},
    \mathcal H_{(3, 5)}\}, \{\mathcal H_{(1, 5)}, \mathcal H_{(2, 4)}\},
    \{\mathcal H_{(2, 3)}, \mathcal H_{(4, 5)}\}\,.
  \end{equation}

  Moreover the over-relaxation updates based on $SU(2)$ projections do
  not need to generate any random numbers, something that might be
  convenient for parallelization.

\item[$U(N)$ over-relaxation] The vectorization of the
  $U(N)$ over-relaxation updates basically requires writing an SVD
  routine that makes use of the vector/multi-core structure of the
  CPU. Some \texttt{LAPACK} implementations have support for
  multi-threading, and represent an alternative way to profit from the
  multi-core structure of the CPU.
\end{description}

%%% Local Variables:
%%% mode: latex
%%% TeX-master: "paper"
%%% End:

\section{Numerical comparison of algorithms}

Since the purpose of any simulation is to compute the expectation
values of some observables with the highest possible accuracy, the
merit of an algorithm has to be evaluated by comparing the
uncertainties of some observables per unit of CPU time. There are only
two factors that have to be taken into account in our particular
case: the CPU time per update sweep, and the autocorrelations of the
measurements. 

After having introduced the auxiliary variables, and transformed the
action in a linear function of the links, one can use different
updating methods: heatbath (HB) and over-relaxation (OR2) based in the
projection onto $SU(2)$ subgroups and the $U(N)$ overrelaxation (see
Algorithm~\ref{alg}).  
Since all three sweep methods have to compute the same
staples and generate the same auxiliary variables Amdahl's law puts a
limit on the theoretical improvement that one sweep method can have
over the others. 

In principle OR2 sweeps are very
simple and do not require to evaluate any mathematical function like
$\cos,\log,\dots$, but we have observed that the CPU time for a HB
sweep is essentially the same as the CPU time required for an OR2 sweep. 

The case of ORN is more difficult to evaluate, and depends crucially
on the implementation of the SVD decomposition. With the standard
\texttt{BLAS/LAPACK} implementation, the time per sweep is roughly the
same as 
the time required for an HB/OR2 sweep (faster for
$N>300$, slightly slower for small values of $N$). But we have
observed that some \texttt{LAPACK} implementations of the SVD
decomposition (like \texttt{intel MKL}) run up $\times 10$
faster for large $N$. This is mainly due to the fact that the
prefetching of data and block decomposition used by many
\texttt{BLAS/LAPACK} efficient implementations actually sustain
the manifest $\mathcal O(N^3)$ scaling of the algorithm up to very
large values of $N$ (when the matrices are actually several hundreds of
MB in size).

In any case all these update methods scale like $\mathcal O(N^3)$, and
hereafter we will assume that the three updating methods (HB/OR2/ORN)
have the same cost per sweep. This reduces the question of the most
efficient algorithm to the one that decorrelates measurements
faster. Nevertheless the reader should keep in mind that
if an efficient implementation of \texttt{BLAS/LAPACK} is available
for the architecture they are running, there might be a saving in CPU
time. 

To make the comparisons we have chosen different observables. Firstly
Wilson loops of different sizes; $W(1,1)$, $W(3,3)$ and
$W(5,5)$. These are the classic observables that have been used in the
past for several studies.

Secondly, smeared 
quantities that are known to have large autocorrelation times. In
particular quantities derived from the gradient
flow have been used to test how the slow modes of algorithms
decorrelate~\cite{Luscher:2011kk}. We will consider the
renormalized coupling $\lambda_{TGF}$~\cite{Ramos:2014kla} 
as defined for the TEK model in
Ref.~\cite{Perez:2014isa}. We will tune $b$ for
various values of $N$, such that $\lambda_{TGF}\simeq 23$, and hence
the physical volume ($a\sqrt{N}$) is kept approximately constant. This
will allow us 
to study how the autocorrelation time of a physical quantity scales as
we approach the continuum limit.

Measuring flow quantities is very expensive for the case
of the TEK model. Compared with a typical update, the integration of
the flow equations can consume $\mathcal O(10^5)$ times more
computer time. This is easy to understand if one takes into account that
the measurement involves the computation of
$\mathcal O(10^4)$ times the exponential of a matrix. Usually it is
not a huge problem to measure these observables with a high accuracy,
since the observables have a very small variance and one can easily 
run many parallel replicas. But in order to measure
autocorrelation times we actually need to collect at least $\mathcal
O(100\tau_{\rm int})$ measurements on the same Monte Carlo
chain. Since $\tau_{\rm int}$ can easily go up to 200, this task is
actually very difficult. We address this difficulty in two ways. First
we measure the flow quantities frequently enough to actually measure
the correlation between the data, but not too frequently. We aim at
measuring integrated autocorrelation times of the order of 10. The
values that we will present for $\tau_{\rm int}$ are then computed by
scaling with the number of updates in between measurements. Second
we use a large step size to integrate the flow equations. We use the
third order Runge-Kuta integrator described in the appendix
of~\cite{Luscher:2010iy}, with a step size of $\epsilon = 0.05$. 
We explicitly checked the size of the systematic error in the measured coupling
due to the numerical integration on several configurations for each value of $N$.
In all cases the size of this relative error on the measured coupling was below $0.005 \%$,
moreover this error decreased with increasing $N$, so the systematic error 
due to the large integration step size is negligible 
compared to our $\mathcal{O}(0.5\%)$ statistical errors.

\subsection{Comparison of link update algorithms and scaling}

Our Monte Carlo simulations produce correlated estimates of the
observables of interest. The correlation between measurements can be
quantified by $\tau_{\rm int}$. We follow the notation and conventions
of~\cite{Wolff:2003sm} and define the autocorrelation
function of the observable $O$ as
\begin{equation}
  \Gamma_O(t) = \langle (O(t)-\overline O)(O(0)-\overline O)\rangle\,,
\end{equation}
were the simulation time $t$ labels the measurements and $\overline O$ is
the average value of our observable of interest, we have
\begin{equation}
  \label{eq:tauint}
  \tau_{\rm int} = \frac{1}{2} + \sum_{t=1}^\infty
  \frac{\Gamma_O(t)}{\Gamma_O(0)}\,. 
\end{equation}
In practical situations the sum Eq.~(\ref{eq:tauint}) has to be
truncated after a finite number of terms $W$ that defines our
\emph{summation window}. Since the
autocorrelation function $\Gamma_O(t)$ decays 
exponentially fast for large $t$, estimates of $\tau_{\rm int}$ 
will be accurate as long as $W$ is large compared with the
slowest mode of the Markov operator. In this work we find it is sufficient to simply
truncate the sum when the estimate of $\tau_{\rm int}$ shows a plateau
in $W$ (see figures below), but other situations may require more
sophisticated techniques (see for example~\cite{Schaefer:2010hu}). 

\begin{figure}
\begin{center}
\includegraphics[width=70mm]{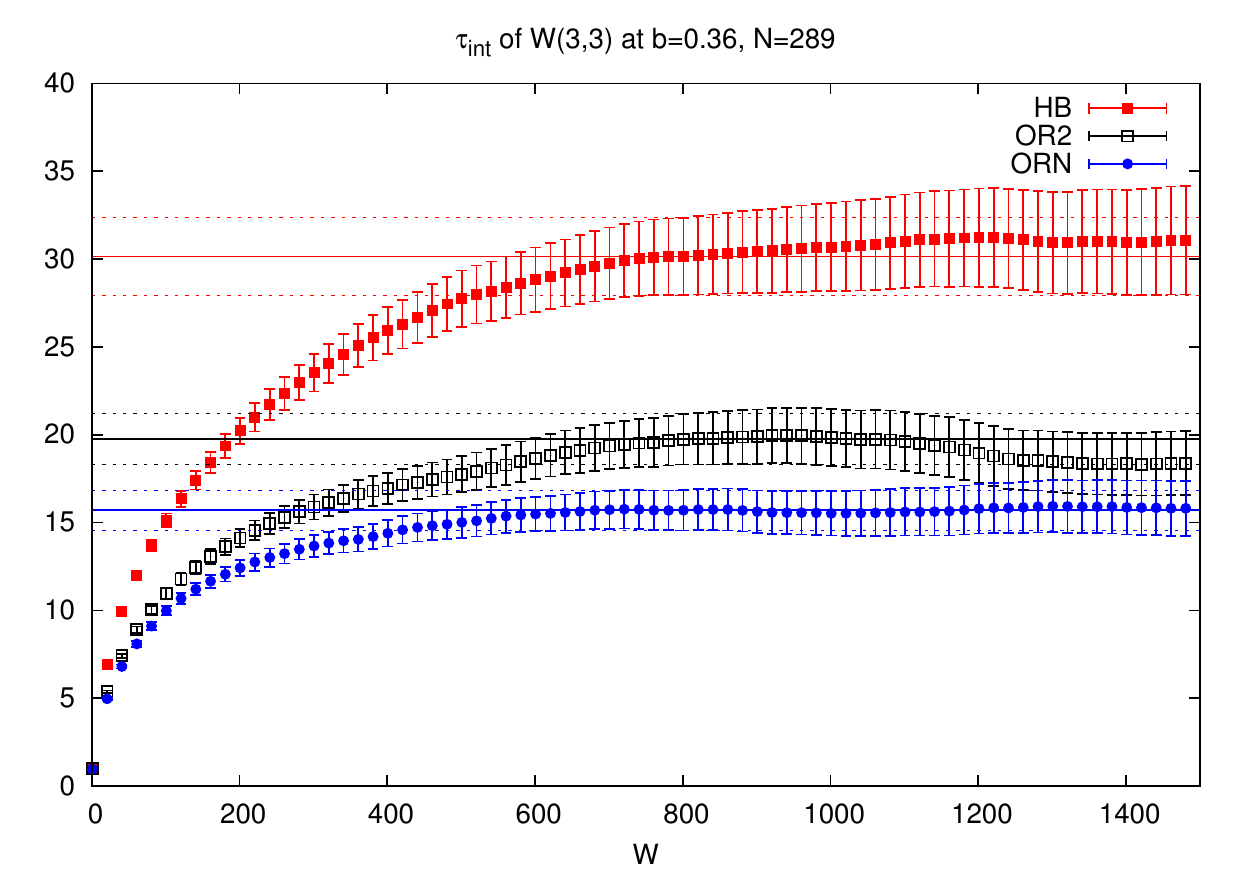} \includegraphics[width=70mm]{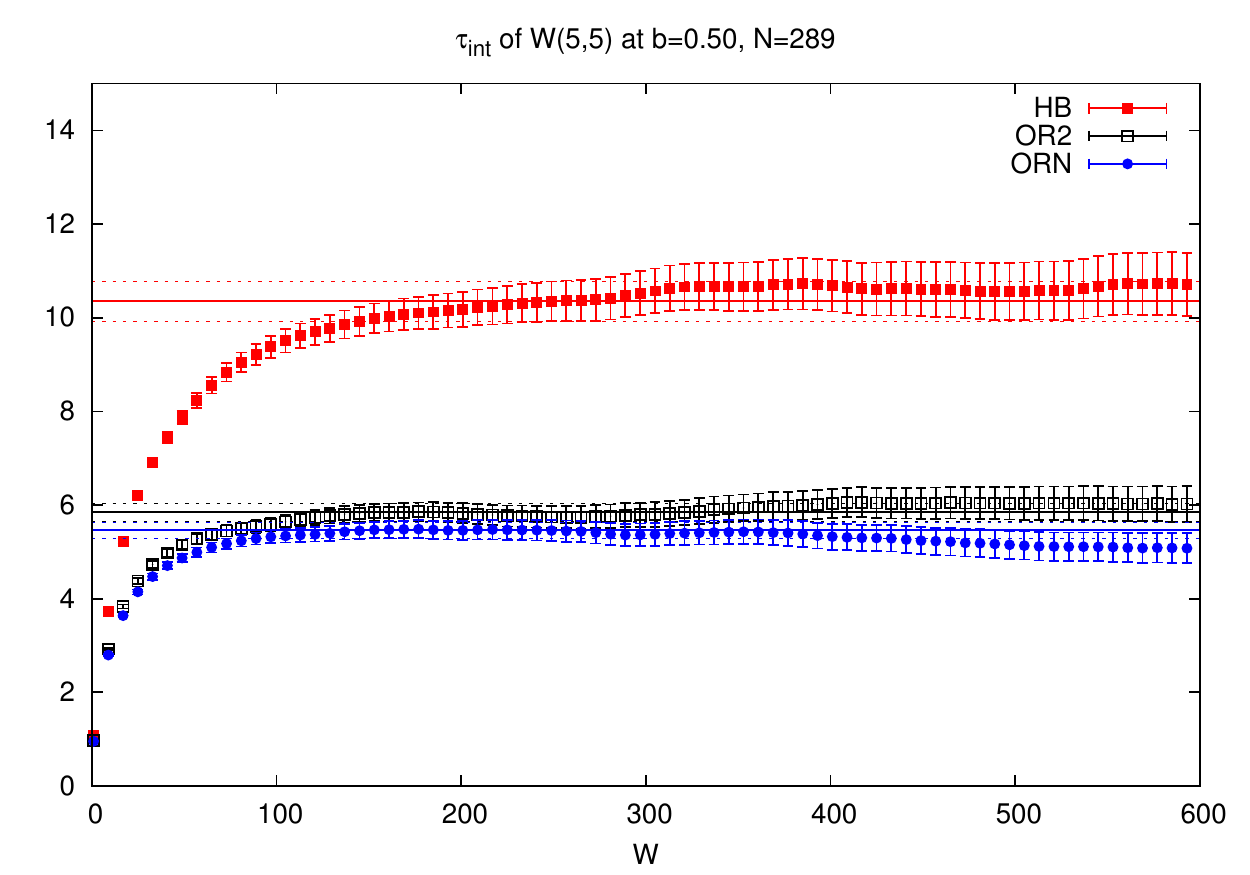}
\end{center}
\caption{\label{fig:wloop}Integrated autocorrelation time 
  estimates as a function of the window size (see Eq.~\ref{eq:tauint})
  for the 3x3 Wilson Loop at $N=289,b=0.36$ (left), and the 5x5 Wilson
  Loop at $N=289,b=0.50$ (right). Different symbols correspond to
  different updates (HB/OR2/ORN). We take $\tau_{int}$ as the plateau
  value.} 
\end{figure}

We now move on to comparing the three different update algorithms, HB,
OR2 and ORN. In Table~\ref{tab:wloop49} we
show the results from the three update methods of Wilson loop
measurements at $N=49$ with $2 \times 10^6$ updates for each.
Similarly Table~\ref{tab:wloop} and Figure~\ref{fig:wloop} show Wilson loop
measurements at $N=289$ with $6 \times 10^5$ updates for each.
All the update methods give observables which are in agreement within errors,
and the OR updates result in observables with about half the
integrated autocorrelation time of the HB update. However, despite the
increased computational complexity of the ORN method compared to the
OR2 method, it does not result in a significantly smaller integrated
autocorrelation time. 

\begin{table}[htb]
\begin{center}
\begin{tabular}{c|c|c}
      $b=0.36$ & $\left<W(1,1)\right>$ & $\tau_{int}$ \\ \hline 
HB & 0.55690(4) & 14.9(3) \\
OR2 & 0.55694(3) & 8.7(2) \\
ORN & 0.55698(3) & 7.2(1) \\
  \end{tabular}
  \end{center} 
 \caption{Wilson Loop average values and integrated correlation times
   at $N=49,k=1$ for the three different update methods.} 
\label{tab:wloop49}
\end{table}

\begin{table}[htb]
\begin{center}
\begin{tabular}{c|c|c}
      $b=0.36$ & $\left<W(3,3)\right>$ & $\tau_{int}$ \\ \hline 
      HB   & 0.032373(15) & 30(2) \\
      OR2   & 0.032401(13) & 20(2)\\
      ORN   & 0.032390(11) & 16(1)\\
  \end{tabular}
  \quad
\begin{tabular}{c|c|c}
      $b=0.50$ & $\left<W(5,5)\right>$ & $\tau_{int}$ \\ \hline 
      HB   & 0.0424222(104) & 10.4(4) \\
      OR2   & 0.0424268(78) & 6.0(4) \\
      ORN   & 0.0424213(76) & 5.5(2) \\
  \end{tabular}

  \end{center} 
 \caption{Wilson Loop average values and integrated correlation times
   at $N=289, k=5$ for the three different update methods.} 
\label{tab:wloop}
\end{table}

\begin{figure}
\begin{center}
\includegraphics[angle=0,width=9cm]{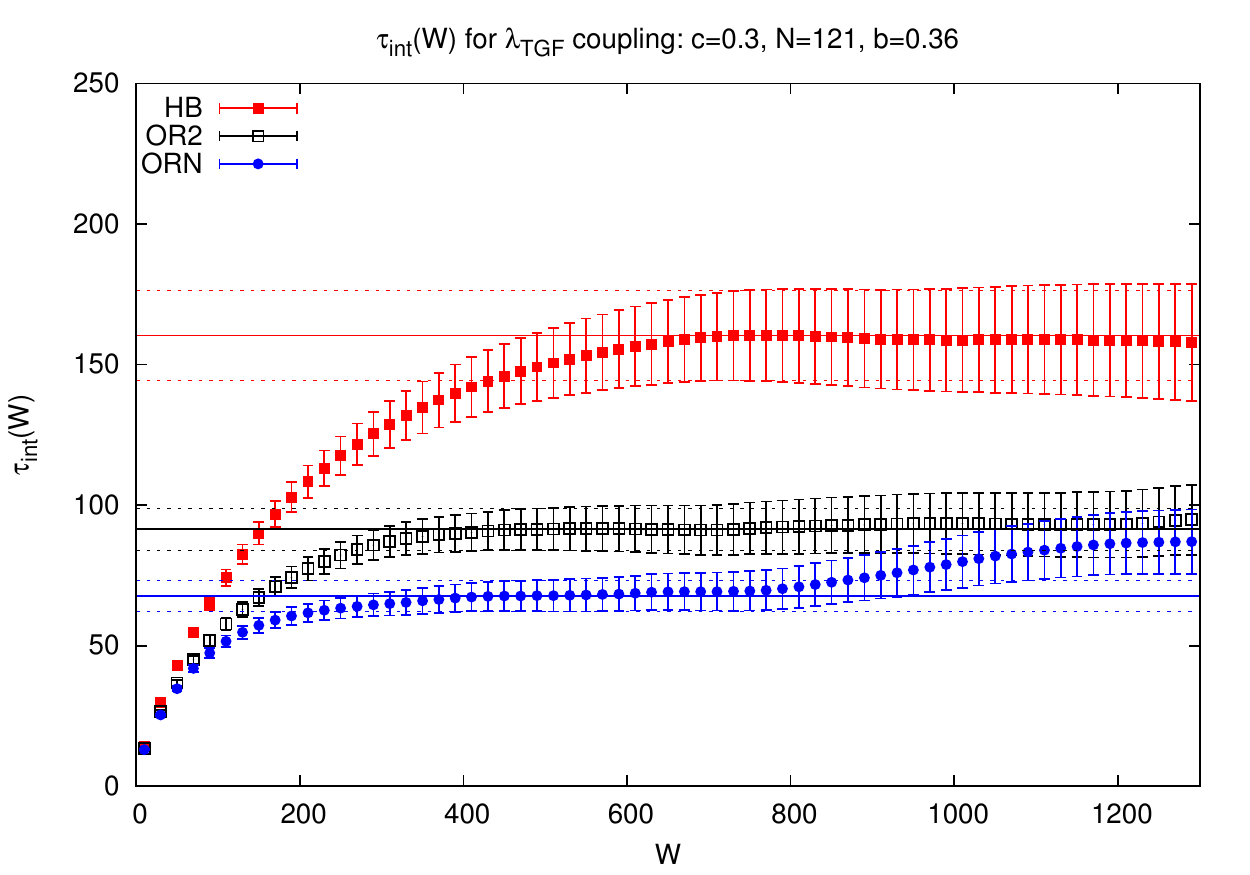}
\caption{Integrated autocorrelation time 
  estimates as a function of the window size (see Eq.~\ref{eq:tauint})
  for the Twisted
  Gradient Flow coupling vs $W$, for $\lambda_{TGF}\simeq23$ at
  $N=121$, $b=0.36$, $c=0.30$. OR updates are significantly better
  than HB, with no significant difference between OR2 and ORN. We take
  $\tau_{int}$ as the plateau value.} 
\label{fig:tau121}
\end{center}
\end{figure}

\begin{figure}
\begin{center}
\includegraphics[angle=0,width=9cm]{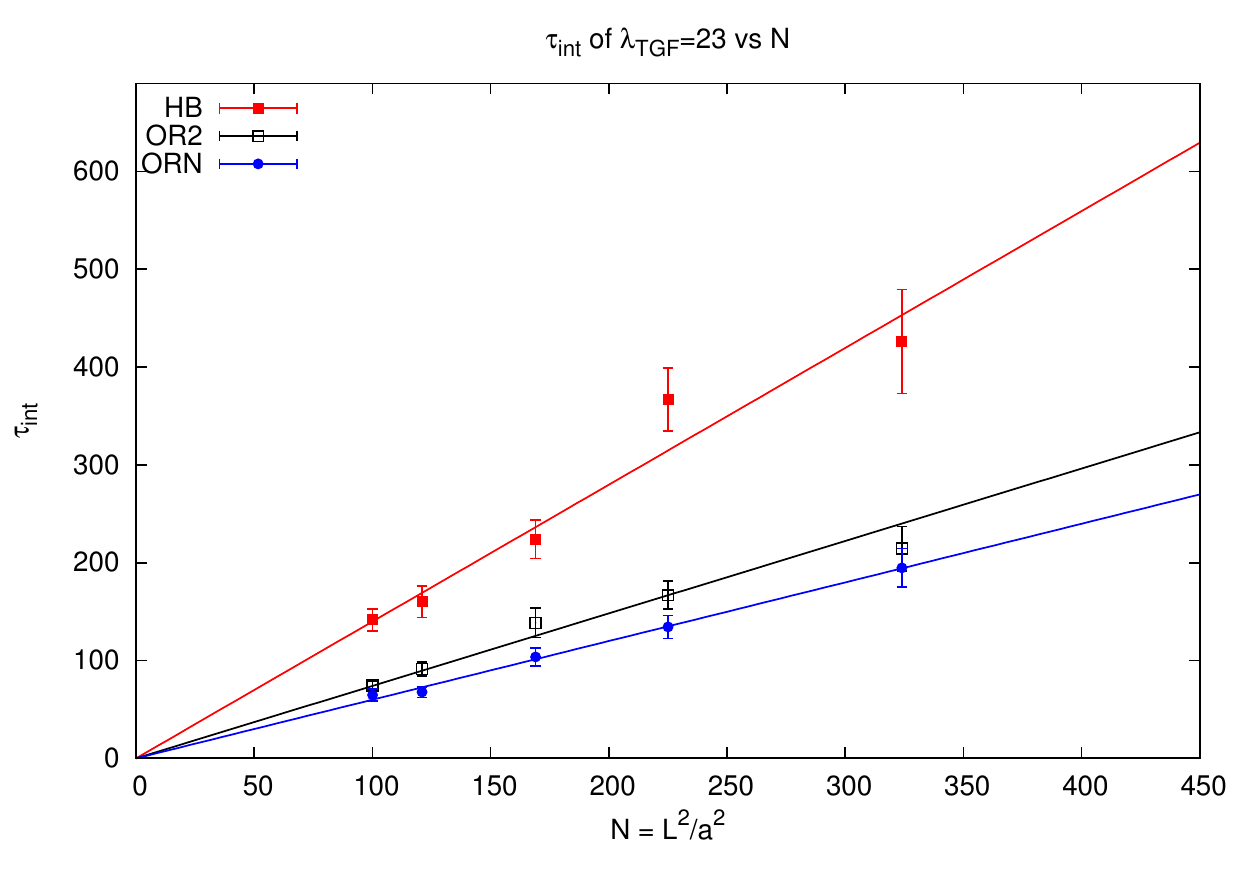}
\caption{ORN Integrated autocorrelation time $\tau_{int}$ vs
  $N=(L/a)^2$, at fixed physical volume, $\lambda_{TGF}\simeq23$. All
  three updates appear to scale with the same critical exponent
  $\tau_{int}\sim a^{-2}$} 
\label{fig:tauN}
\end{center}
\end{figure}

Figure~\ref{fig:tau121} shows the integrated autocorrelation time of
$\lambda_{TGF}$ for $N=121$ for the three update algorithms. We see
the same results as for the Wilson loop observables; OR updates have
around half the $\tau_{int}$ of HB updates, with little difference
between OR2 and ORN. Since we have tuned the physical volume to be
constant, we can repeat these measurements at different values of $N$
to determine how the integrated correlation time scales as a function
of the lattice spacing $a$, using the relation $\sqrt{N}=L/a$. The
results are listed in Table~\ref{tab:coupling} for the three update methods, 
and $\tau_{int}$ as a function of $a^{-2}$ is shown in
Figure~\ref{fig:tauN}.

\begin{table}[htb]
\begin{center}
\begin{tabular}{c|c|c|c|c|c|c|c|c}
   \multicolumn{3}{c|}{} & \multicolumn{2}{|c|}{HB} & \multicolumn{2}{|c|}{OR2}  &\multicolumn{2}{|c}{ORN}  \\ \hline 
$N$ & $b$ & $k$ & $\lambda_{TGF}$ & $\tau_{int}$ & $\lambda_{TGF}$ & $\tau_{int}$ & $\lambda_{TGF}$ & $\tau_{int}$ \\ \hline 
100 & 0.358 & 3 & 22.3(1) & 142(11) & 22.3(1) & 74(4) & 22.0(1) & 65(6) \\
121 & 0.360 & 3 & 22.8(2) & 160(16) & 22.8(1) & 91(8) & 22.6(1) & 68(6) \\
169 & 0.365 & 4 & 24.3(2) & 224(20) & 24.2(2) & 139(15) & 24.3(2) & 104(9) \\
225 & 0.372 & 4 & 21.8(1) & 367(32) & 21.9(1) & 167(14) & 21.8(1) & 134(12) \\
324 & 0.375 & 5 & 24.2(2) & 426(53) & 24.1(2) & 214(23) & 24.0(2) & 195(20) \\
  \end{tabular}
  \end{center}
 \caption{$\lambda_{TGF}\simeq23$ average values and integrated
   correlation times at each $N$ for the three different update
   methods.} 
\label{tab:coupling}
\end{table}

The generic expectation in normal lattice simulations is for a local
update to behave like a random walk, and so for $\tau_{int}$ to scale
like $a^{-2}$. A priori it is not clear that
this should also apply to the TEK model, since in this case a
``local'' update of a single gauge link actually involves all the
degrees of freedom in the lattice. It turns out however that all three
update methods exhibit the same critical scaling as one would expect
for a local algorithm, namely $\tau_{int} \sim a^{-2}$, as can be seen
in Figure~\ref{fig:tauN}.

%%% Local Variables:
%%% mode: latex
%%% TeX-master: "paper"
%%% End:

\section{Conclusions}

We have studied several simulation algorithms for one-site lattice
models. In particular we have focused on  the TEK model, which is relevant
in the context of the large $N$ expansion of gauge theories and has
recently been used to compute many interesting properties of
$SU(\infty)$ gauge
theories~\cite{GonzalezArroyo:2012fx,Perez:2014isa,Perez:2014gqa}. 

Following Fabricius and Haan~\cite{Fabricius:1984wp} we introduce auxiliary variables
to make the action a linear function of the links, and study several
link-update algorithms. Up to now all authors included a combinaton of
heat-bath steps  
and overrelaxation steps in their algorithms. However, we show that this 
is not necessary, since once the auxiliary variables are updated,
overrelaxation alone suffices to make the algorithm ergodic. This is in  
contrast with the usual lattice gauge theory, where over-relaxation sweeps 
produce microcanonical movements and are therefore not ergodic. 

Regarding the over-relaxation updates, we study two kinds. First the
one based in the projection over $SU(2)$ subgroups (OR2). Second we
study the OR method over the whole  group as proposed
in~\cite{Kiskis:2003rd,deForcrand:2005xr} (ORN). Indeed, we realize 
that for one-site models the algorithm does not change the value of
the action, making a Metropolis accept/reject step unnecessary. This 
is due to the equivalence of U(N) and SU(N) groups for these models. 

Finally, we perform a performance comparison between the different
alternatives.   We show that, at different values of $N$ for different
observables, overrelaxation sweeps decorrelate faster than heatbath sweeps ($\tau_{\rm int}^{\rm
  OR}/\tau_{\rm int}^{\rm HB} \simeq 1/2$). We see no big differences
in terms of autocorrelation times between the two possible
overrelaxation methods (OR2 and ORN). Each algorithm has his own benefits.
OR2 is simple and easy to vectorize. On the other hand ORN might profit from a highly optimized
routine for matrix operations, in particular the SVD decomposition. We
conclude by studying the performance for computing a renormalized
quantity in the continuum limit, the twisted gradient flow
renormalized coupling. We see that the three tested link-update
algorithms scale like $a^{-2}$. Hence,   none of them
has a better critical exponent.

%%% Local Variables:
%%% mode: latex
%%% TeX-master: "paper"
%%% End:

\section*{Acknowledgments}

L.K. and A.R. want to thank A. Patella for many interesting
discussions. A.R. wants also to thank I. Campos for providing
computer time in the IFCA clusters.
We acknowledge
financial support from the MCINN 
grants FPA2012-31686 and FPA2012-31880, 
and the Spanish MINECO’s ``Centro
de Excelencia Severo Ochoa'' Programme under grant
SEV-2012-0249. M. O. is supported by the Japanese MEXT grant No
26400249. Calculations have been done on Hitachi SR16000 supercomputer
both at High Energy Accelerator Research Organization(KEK) and YITP in
Kyoto University, the ALTAMIRA clusters at IFCA, and the HPC-clusters
at IFT. Work at KEK is supported by the Large Scale Simulation Program
No.14/15-03.  

\clearpage

\appendix

\section{$SU(N)$ overrelaxation}
\label{ap:orsun}

Let us recall that  the  overrelaxation  transformation takes the form:
\begin{equation}
U^{\rm new}_{\alpha} = W_{\alpha} U_{\alpha}^{\dagger} W_{\alpha}\,.
\end{equation}
For the $U(N)$ case one takes $W_{\alpha}=V_{\alpha}
X^{\dagger}_{\alpha}$, where $X$ and $V$ are $U(N)$ matrices obtained
from the SVD decomposition of the staple:
\begin{equation}
H_{\alpha}=X_{\alpha} \sigma V^{\dagger}_{\alpha}\,.
\end{equation}
However, this transformation is not valid for $SU(N)$ since 
\begin{equation}
\det[V_{\alpha} X^{\dagger}_{\alpha}] = e^{-i\Phi} \neq 1\,.
\end{equation}
Following Ref.~\cite{Kiskis:2003rd},  we will propose the $SU(N)$ matrix 
\begin{equation}
W_{\alpha}'=V_{\alpha} D_{\alpha} X^{\dagger}_{\alpha}\,,  
\end{equation}
where $D={\rm diag}(e^{i\theta_1},\dots,e^{i\theta_N})$, with the condition
\begin{equation}
\label{constraint}
\sum_i \theta_i = \Phi \mod 2\pi
\end{equation}

The simplest way to implement the constraint would be  to take 
$\theta_i = \Phi/N$, as done in Ref.~\cite{Kiskis:2003rd}. This 
is then followed by an accept/reject Metropolis step as explained
earlier.  We find this works reasonably well in
practice, with an acceptance rate of $\sim85\%$ at $b=0.36$ for a
range of values of $N$ from $81$ to $1369$, with no noticeable
dependence on $N$, and with higher acceptance rates at weaker values of
the coupling. 

As suggested in Ref.~\cite{deForcrand:2005xr}, the acceptance rates 
can be improved by tuning the angles $\theta_i$ to minimize the quantity ${\rm ReTr}
\left[ W_\alpha H_\alpha \right]$, i.e. choose $\theta_i$ such that 
\begin{equation}
\min_{\theta_i}\left\{ \sum_i \sigma_i \cos(\theta_i) \right\}, \quad
\sum_i \theta_i = \Phi \mod 2\pi\,. 
\end{equation}
Since the minimization has to be done preserving the constraint
Eq.~(\ref{constraint}), the best way is to add a lagrange multiplier
$\lambda$ enforcing the condition in the minimization function:
\begin{equation}
\min_{\theta_i}\left\{ \sum_i \sigma_i \cos(\theta_i) + \lambda\left(\sum_i \theta_i - \Phi\right) \right\}\,. 
\end{equation}
The equations for minimum lead to the solution 
\begin{equation}
  \theta_i = \arcsin\frac{\lambda}{\sigma_i} 
\end{equation}
where the quantity $\lambda$ satisfies the equation
\begin{equation}
\sum_i \arcsin\frac{\lambda}{\sigma_i} = \Phi\,.
\end{equation}
The latter is a single transcendental equation, which  
can be solved by any standard technique, like Newton--Raphson. 
We have observed that this procedure converges
very fast, and in fact increases the acceptance rate to about $\sim
98\%$. 

%%% Local Variables:
%%% mode: latex
%%% TeX-master: "paper"
%%% End:

\bibliography{campos}

\end{document}